\documentclass[floatfix,twocolumn,aps,prd,showpacs,amsmath,amssymb]{revtex4-1}
\usepackage[latin1]{inputenc}
\usepackage[dvips]{graphicx}

\usepackage{dcolumn}
\usepackage{bm}
\usepackage{dsfont}
\usepackage{color}
\usepackage{ulem} 
\usepackage{url}
\usepackage[colorlinks=true ,citecolor=blue]{hyperref}
\usepackage{amsmath,amssymb}

\newcommand{\PR}[1]{\ensuremath{\left[#1\right]}}
\newcommand{\PC}[1]{\ensuremath{\left(#1\right)}}
\newcommand{\chav}[1]{\ensuremath{\left\{#1\right\}}}

\begin{document}

\title{Spin $g$-factor due to electronic interactions in graphene}

\author{Nat\'alia Menezes$^{1,2}$, Van S\'ergio Alves$^{2,3}$,  E. C. Marino$^{3}$, Leonardo Nascimento$^{2,4}$,  Leandro O. Nascimento$^{1,3}$, C. Morais Smith$^{1}$ }
\affiliation{$^1$Institute for Theoretical Physics, Center for Extreme Matter and Emergent Phenomena, Utrecht University, Princetonplein 5, 3584 CC Utrecht, The Netherlands \\
$^2$Faculdade de F\'\i sica, Universidade Federal do Par\'a, Avenida Augusto Correa 01, 66075-110, Bel\'em, Par\'a,  Brazil \\
$^3$Instituto de F\'\i sica, Universidade Federal do Rio de Janeiro, C.P.68528, Rio de Janeiro RJ, 21941-972, Brazil \\
$^4$Instituto Federal do Par\'a, 66093-020, Bel\'em, Par\'a, Brazil }

\date{\today}

\begin{abstract}
The gyromagnetic factor is an important physical quantity relating the magnetic-dipole moment of a particle to its spin. The electron spin $g$-factor {\it in vacuo} is one of the best model-based theoretical predictions ever made, showing agreement with the measured value up to ten parts per trillion \cite{Schwinger48,expvan,Hanneke,Kinoshita}. However, for electrons in a material the $g$-factor is modified with respect to its value {\it in vacuo} because of environment interactions.
Here, we show how interaction effects lead to the spin $g$-factor correction in graphene by considering the full electromagnetic interaction in the framework of pseudo-QED \cite{Kovner,Marino}. We compare our theoretical prediction with experiments performed on graphene deposited on SiO$_2$ and SiC, and find a very good agreement between them. 
\end{abstract}

\maketitle

\section{Introduction}

 The electron dispersion relation in solid-state materials strongly depends  on the crystal-lattice geometry. In the case of graphene, the honeycomb lattice leads to a zero-mass relativistic-like dispersion $E_{\pm}(\textbf{k})\approx \pm v_F |\textbf{k}|$.  This characteristic allows us to relate the electrons in graphene to free Dirac massless particles in (2+1) dimensions (D) \cite{Geim1}. However, the fact that the photons propagate with the speed of light $c$ and the electrons move with the Fermi velocity $v_{F}\simeq c/300$ \cite{Castro} has important consequences upon the physical properties of the system.

Until a few years ago, graphene was believed to be an effectively noninteracting system. The recent measurement of the fractional quantum Hall effect \cite{Eva,Kim1,Kim2}, which is a typical feature of strongly correlated systems, however, has changed this paradigm. The relevance of interactions in graphene was further confirmed by the experimental observation of the renormalization of the Fermi velocity \cite{Luican, Elias, Stroscio2}, as had been theoretically predicted earlier \cite{Maria0,Maria,Maria2,Foster,Maria3}. More recently, higher-order loop calculations have been performed \cite{dasSarma}. However, most of the theories found in the literature consider only static interactions because $v_F\ll c$. Dynamical effects, nevertheless, have proven to be important in some cases, by generating novel quantum topological states that would not arise in the static limit \cite{us}.

Even though the electrons in graphene are constrained to move on a plane, the electromagnetic field through which they interact spreads in 3D. Integrating away out-of-the-plane photons, one obtains an effective interaction that is non-local in space and time. In spite of being fully 2D, it conveys all properties of the genuine 3D electromagnetic interaction. This interaction has been called pseudo-QED (PQED) because it involves pseudo-differential operators, but sometimes the name reduced QED is also used in the literature \cite{Miransky}. It has been shown to respect causality  \cite{marinorubens}, scale invariance, the Huygens principle, and unitarity \cite{us2}, apart from exhibiting an $1/R$ static Coulomb potential. Actually, the propagator in pseudo-QED in coordinate space coincides with the one of QED$_{2+1}$ in momentum space \cite{marinorubens}, and these two theories are dual to each other \cite{us2}.

Motivated by the relevance of electron-electron interactions in graphene, and by the fact that the Fermi velocity is much different than the speed of light, we investigate in this paper the spin gyromagnetic factor in graphene by using the {\it anisotropic} PQED, which contains a term that breaks Lorentz invariance in the quantum-field-theory formalism. Since the non-local gauge field produces the full electromagnetic interaction, independently on whether the matter is relativistic or not \cite{Marino}, we can easily include the Lorentz violating term in the matter field. We then calculate the spin $g$-factor ($g_s$) and show that it compares very well to the experimental data available in the literature \cite{Kurganova,Stroscio}. Our results set the importance of interactions in determining the $g$-factor in graphene in particular, and 2D relativistic condensed-matter systems in general.

This paper is divided as follows: In Sec.~II, we present the anisotropic version of the PQED theory together with the rules needed to compute Feynman diagrams. In Sec.~III, we discuss how the tree-level vertex diagram leads to corrections to the bare $g$-factor $g_s=2$ due to interaction effects. The detailed calculation of the correction is performed in Sec.~IV, and a comparison of our results with experiments on the spin $g$-factor of graphene deposited on different substrates is shown in Sec.~V. We present our conclusions in Sec.~VI.

\section{Anisotropic pseudo-QED}

The anisotropic version of the PQED is given by the Lagrangian 
\begin{eqnarray}
\mathcal{L} &=& -\frac{1}{2} F_{\mu\nu}\frac{1}{\sqrt{\Box}}F^{\mu\nu} + \bar{\psi}_{\kappa}(i \gamma^{0}\partial_{0} +i v_{F}\gamma^{i} \partial_{i} - \Delta)\psi_{\kappa} \nonumber \\
&-& e\bar{\psi}_{\kappa}\PC{\gamma^{0}A_{0}+\frac{v_F}{c}\gamma^{i}A_{i}} \psi_{\kappa} + \frac{\zeta}{2} A_{\mu} \frac{\partial^{\mu}\partial^{\nu}}{\sqrt{\Box}}A_{\nu}, \label{anisotropic}
\end{eqnarray}
where $F^{\mu\nu}$ is the usual field-intensity tensor of the U(1) gauge field $A_\mu$, which intermediates the electromagnetic interaction in 2D (pseudo electromagnetic field), $\Box$ is the d'Alembertian, and $\bar{\psi}_{\kappa} = \psi^\dagger_{\kappa} \gamma_0$ is the Dirac spinor, with $\kappa$ representing a sum over valleys ($K$ and $K'$). Here, we use the Dirac basis for the $\gamma$-matrices and consider a $4\times 4$-spinor representation $ \psi^\dagger_{\kappa} = (\psi^{\star}_{A \downarrow}, \psi^{\star}_{B \downarrow}, \psi^{\star}_{A \uparrow}, \psi^{\star}_{B \uparrow})_{\kappa}$, with $A$ and $B$ denoting the sublattices in graphene and $\uparrow,\downarrow$ the different spins. The parameter $\zeta$ is the gauge fixing (we adopt Feynman's gauge $\zeta =1$), and $\Delta$ is a gap that may occur due to a sublattice asymmetry in case of graphene deposited on substrates (which also acts as an infrared regularization parameter) \cite{Ando}.  

The Feynman's rules of the model yield the fermion propagator $S_F$, 
\begin{equation}
S_{F} (\bar p) = i \frac{\gamma^\mu \bar p_\mu +\Delta}{\bar p^2-\Delta^2}, \label{01}
\end{equation}
where for Dirac matrices $\gamma^\mu=(\gamma^0,\gamma^i)$, ${\bar p}^{\mu}=(p_0,v_F \textbf{p})$ and $\bar p^2=p_0^2-v_F^2 \textbf{p}^2$. The photon propagator reads
\begin{equation}
G_{\mu\nu} (p) = \frac{-ic}{2\varepsilon\sqrt{p^2}}\PR{g_{\mu\nu}-\PC{1-\frac{1}{\zeta}}\frac{p_{\mu}p_{\nu}}{p^2}}, \label{02}
\end{equation}
where $p_{\mu}$ is the four-momentum given by $p_{\mu}=(p_0,c\textbf{p})$, $ p^2=p_0^2-c^2 \textbf{p}^2$, $g_{\mu\nu}=(1,-1,-1)$, and $\varepsilon$ is the electric permitivity.
The interaction vertex is given by 
\begin{equation}
\Gamma^{\mu}_{0}=-ie\PC{ \gamma^{0}, \beta\,\gamma^{j}}, \label{03}
\end{equation} 
where $\beta \equiv v_{F}/c$.

The pole of the fermion propagator provides the energy dispersion relation $p_{0} = E(\textbf{p})=\pm\sqrt{v_F^2 \textbf{p}^2+\Delta^2}$. When $\Delta =0$, we reproduce the tight-binding result for monolayer graphene.

The first term present in the Maxwell Lagrangian in Eq.~(\ref{anisotropic}) is non-local and renders the canonical dimension of the gauge field equal to one, in units of mass. The same holds for the Dirac field.  Therefore, the coupling constant $e$ is dimensionless in the 2+1 space-time, and the theory is renormalizable, analogously to QED$_{3+1}$. Here, we will calculate the one-loop correction to the vertex diagram using the dimensional regularization procedure as a way to obtain finite Feynman amplitudes, which do not depend on the regulator \cite{Hooft}.

\section{The (2+1)D vertex function}
\begin{figure}[h]
\centering
\includegraphics[scale=0.20]{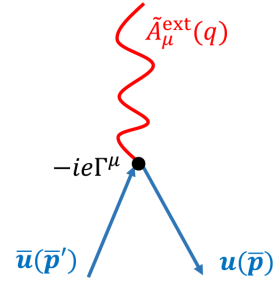}
\caption{Tree-level diagram}
\label{divergences}
\end{figure}
We start by analyzing the $\mathcal{S}$-matrix element $\mathcal{M}$ for the scattering from an external field, represented by the tree-level diagram in Fig.~\ref{divergences}, and written down as \cite{Peskin}
\begin{eqnarray}
i\mathcal{M} (2\pi)\delta(p'^{0}-p^{0})=-ie\bar{u}(\bar{p}')\Gamma^{\mu}u(\bar{p})\cdot \tilde{A}_{\mu}^{{\rm ext}}(p'-p),\label{1}
\end{eqnarray}
where $\bar{u}$ and $u$ are normalized solutions of the free Dirac equation \cite{Drell}, and $\tilde{A}_{\mu}^{{\rm ext}}(\bar{p})$ is the Fourier transform of $A_{\mu}^{{\rm ext}}(x)$, which is a classical external potential. By splitting the different vertex contributions in Eq.~(\ref{1}), we obtain 
\begin{eqnarray}
&&i\mathcal{M} (2\pi)\delta(p'^{0}-p^{0}) =\nonumber \\
&& -ie\bar{u}(\bar{p}')\Gamma^{0}u(\bar{p}) \tilde{\phi}^{{\rm ext}}(q)+ie\beta\bar{u}(\bar{p}')\Gamma u(\bar{p}) \cdot \tilde{\textbf{A}}^{{\rm ext}}(q),
\label{2}
\end{eqnarray}
with $p'-p=q$. Here, $\tilde{\phi}^{{\rm ext}}$ and $\tilde{\textbf{A}}^{{\rm ext}}$ are the scalar and the vector potential, respectively. Lorentz invariance allows us to write the vertex $\Gamma^{\mu}$ as
\begin{eqnarray}
\Gamma^{\mu} = C_1\gamma^{\mu} + C_2 (\bar{p}'^{\mu}+\bar{p}^{\mu}) + C_3 (\bar{p}'^{\mu}-\bar{p}^{\mu}),
\label{3}
\end{eqnarray}
where $C_i$'s are scalar functions of the momentum and/or the fermionic mass. By applying the Ward identity $q_{\mu}\Gamma^{\mu}=0$ in Eq.~(\ref{3}), we find that $C_3=0$. Therefore,
\begin{eqnarray}
\Gamma^{\mu} = C_1\gamma^{\mu} + C_2(\bar{p}'^{\mu}+\bar{p}^{\mu}).\label{4}
\end{eqnarray}

Now, using the Gordon identity, we rewrite Eq.~(\ref{4}) as
\begin{eqnarray}
\bar{u}(\bar{p}') \Gamma^{\mu} u(\bar{p}) =\bar{u}(\bar{p}') \PR{\gamma^{\mu}F_{1}(\bar{q}^{2}) + \frac{i\sigma^{\mu\nu}\bar{q}_{\nu}}{2\Delta}F_{2}(\bar{q}^{2}) }u(\bar{p}),\nonumber \\
 \label{5}
\end{eqnarray}
where $F_{1}$ and $F_{2}$ are form factors. At the tree-level diagram, $F_{1}=1$ and $F_{2}=0$. Plugging the above result in Eq.~(\ref{1}), we have
\begin{eqnarray}
&&i\mathcal{M} (2\pi)\delta(q_{0})= -ie \times\nonumber \\
&& \bar{u}(\bar{p}')\PR{\gamma^{\mu}F_{1}(\bar{q}^{2}) + \frac{i\sigma^{\mu\nu}\bar{q}_{\nu}}{2\Delta}F_{2}(\bar{q}^{2}) }u(\bar{p})\cdot \tilde{A}_{\mu}^{{\rm ext}}(q) . \label{6}
\end{eqnarray}
So far, we did not specify the spacetime dimension of the system studied. To understand better the problem in (2+1)D, let us follow the analysis performed in Ref.~\cite{Peskin}, but now for $\mu,\nu=0,1,2$. 

Focusing on the spatial component of the four-vector potential $A_{\mu}^{{\rm ext}}(x)=(0,\textbf{A}^{{\rm ext}}(\textbf{x}))$, or in the Fourier space $\tilde{A}_{\mu}^{{\rm ext}}(\bar{q})=(0,\tilde{\textbf{A}}^{{\rm ext}}(\textbf{q}))$, one obtains
\begin{eqnarray}
&&i\mathcal{M}= +ie \beta \times\nonumber \\
&& \bar{u}(\bar{p}')\PR{\gamma^{i}F_{1}(\bar{q}^{2}) + \frac{i\sigma^{i\nu}\bar{q}_{\nu}}{2\Delta}F_{2}(\bar{q}^{2}) }u(\bar{p})\cdot \tilde{\textbf{A}}^{i}_{{\rm ext}}(\textbf{q}).\label{7}
\end{eqnarray}
By performing a non-relativistic expansion of the spinor, i.e.
$$u(\bar{p})= \left( \begin{array}{c}
\sqrt{\bar{p}\cdot\sigma}\xi  \\
\sqrt{\bar{p}\cdot\bar{\sigma}}\xi  \end{array} \right) \approx \sqrt{\Delta} \left( \begin{array}{c}
(1-v_{F}\boldsymbol{p}\cdot\boldsymbol{\sigma}/2\Delta)\xi  \\
(1+v_{F}\boldsymbol{p}\cdot\boldsymbol{\sigma}/2\Delta)\xi  \end{array} \right) ,$$
with $\xi$ a spinor in the spin space, $\sigma=(\mathds{1},\sigma^{i})$ and $\bar{\sigma}=(\mathds{1},-\sigma^{i})$, the first term in Eq.~(\ref{7}) yields
\begin{eqnarray}
\bar{u}(\bar{p}') \gamma^{i}u(\bar{p}) &=& 2\Delta v_{F} \xi'^{\dagger}\PC{\frac{\boldsymbol{p'}\cdot\boldsymbol{\sigma}}{2\Delta}\sigma^{i}+\sigma^{i}\frac{\boldsymbol{p}\cdot\boldsymbol{\sigma}}{2\Delta} }\xi\nonumber\\
&=&  2\Delta v_{F} \xi'^{\dagger}\PR{\frac{P^{j}\delta^{ji}\mathds{1}-i\varepsilon^{ijk}q^{j}\sigma^{k}}{2\Delta}}\xi,
\label{8}
\end{eqnarray}
where $P^{j}=p'^{j}+p^{j}$. The first term in Eq.~(\ref{8}) is a contribution from the operator $\textbf{p}\cdot\textbf{A}+\textbf{A}\cdot \textbf{p}$, while the second term is the magnetic-moment interaction. Notice that although in a strictly two-dimensional system the momentum $p_{z}=0$ (i.e. $j=1,2$), the set of Pauli matrices encounters the possibility of $k=0,1,2$. Hence, for a non-vanishing magnetic moment interaction, there are two possibilities for the Levi-Civita, $\varepsilon^{120}$ and $\varepsilon^{210}$, which leads to
\begin{eqnarray}
&& i (2\Delta\beta)  \xi'^{\dagger}\PC{\frac{2e}{2\Delta}}\frac{\sigma^{0}}{2}\xi\cdot  (-i\varepsilon^{ij0}v_{F}q^{j}\tilde{\textbf{A}}_{{\rm ext}}^{i}(\textbf{q})) \nonumber \\
&&=- i (2\Delta\beta)  g_{s}\mu_{B}\xi'^{\dagger}\frac{\sigma^{0}}{2}\xi\cdot  (-\nabla_{\perp}\cdot \textbf{A}_{{\rm ext}}(\textbf{x})),\nonumber \\
&&=- i (2\Delta\beta)  g_{s}\mu_{B}S B_{\perp}. \label{g2}
\end{eqnarray}
Here, we used that $q \rightarrow -i\partial$ with $\nabla_{\perp}=(\partial_{y},-\partial_{x})$, $\mu_{B} = e/2\Delta$ is the Bohr magneton, $S$ is the electron's spin, $B_{\perp}$ is a magnetic field perpendicular to the electron's propagation and $g_{s}=2$ (non-interacting case).

Proceeding with a similar analysis for the second term in Eq.~(\ref{7}), we obtain
\begin{eqnarray}
\bar{u}(\bar{p}')\sigma^{i\nu}\bar{q}_{\nu}u(\bar{p}) = 2\Delta \xi'^{\dagger} \varepsilon^{ij0}\sigma_{0}v_{F}q_{j}\xi.\label{9}
\end{eqnarray}
Now, by rewriting the contribution from Eq.~(\ref{9}) as the one in Eq.~(\ref{g2}) and replacing both results together with Eq.~(\ref{8}) into Eq.~(\ref{7}), we obtain
\begin{eqnarray}
&&i\mathcal{M}=i (2\Delta\beta) \xi'^{\dagger}\PC{\frac{ev_{F}P^{i}\mathds{1}}{2\Delta} }F_{1} \xi \cdot\tilde{\textbf{A}}^{i}_{{\rm ext}}(\textbf{q}) -\nonumber \\
&& i (2\Delta\beta)  g_{s}(F_1+F_2)\mu_{B}S B_{\perp}. \label{15} 
\end{eqnarray}

In the second term of Eq.~(\ref{15}), we observe how interaction effects can change the value of the spin $g$-factor, leading to a corrected $g_{s}^{*}$ ($F_1=1$),
\begin{eqnarray}
g_{s}^{*}\equiv 2+ 2F_2 = 2+\mathcal{O}(\alpha). \label{16} 
\end{eqnarray}
In the following section, we calculate the value of this correction, i.e. the form factor $F_{2}$.

\begin{widetext}

\section{Form factor calculation}

\begin{figure}[h!]
\centering
\includegraphics[width=0.15\textwidth]{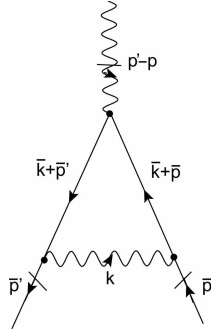}
\caption{One-loop vertex correction.}
\label{vertexgraph}
\end{figure}

Our aim in this section is to compute the one-loop correction to the electron's gyromagnetic factor $g_{s}$ using the anisotropic pseudo-QED. For this, it is only necessary to calculate the finite part of the spatial component of the vertex represented in Fig.~\ref{vertexgraph}. According to Feynman's rules, the vertex diagram is given by

\begin{eqnarray}
i\mathcal{M}= +ie\beta\bar{u} \int \frac{d^{3}k}{(2\pi)^{3}} \chav{\Gamma^{\alpha}_{0}S_{F}(\bar{k}+\bar{p}')\gamma^{i} S_{F}(\bar{k}+\bar{p})\Gamma^{\beta}_{0}G_{\alpha\beta}(k)}u  \tilde{\textbf{A}}^{i}_{{\rm cl}},  \label{S4E1}
\end{eqnarray}
with $\mathcal{M}=\Omega^{i}\tilde{\textbf{A}}^{i}_{{\rm cl}}$, and
\begin{eqnarray}
\Omega^{i}&=& -\frac{ie^{3}v_F}{2\varepsilon}\int \frac{d^{3}k}{(2\pi)^{3}}\bar{u}\chav{\frac{\gamma^{\alpha}\PC{ \displaystyle{\not}\bar{k}+ \displaystyle{\not}\bar{p}' + \Delta}\gamma^{i}\PC{ \displaystyle{\not}\bar{k}+ \displaystyle{\not}\bar{p} + \Delta}\gamma_{\alpha}}{\PR{(\bar{k}+\bar{p}')^2-\Delta^2}\PR{(\bar{k}+\bar{p})^2-\Delta^2}\sqrt{k_{0}^2-c^2\vec{k}^2}}}u. \label{S4E2}
\end{eqnarray}
\end{widetext}

To solve Eq.~(\ref{S4E2}) and find the correction to the bare $g$-factor, first we rewrite the numerator of the integrand by using the properties of gamma matrices and the Dirac equations for $u$ and $\bar{u}$. Then, we parametrize the denominator in order to obtain a single function of the momentum $k$, thus simplifying the integrals. By evaluating the integrals over both $k_0$ and $\textbf{k}$ separately, and focusing on the relevant terms to generate the anomalous gyromagnetic factor (see Appendix for details of the calculations), we find
\begin{equation}
\Omega^{i}_{gy}=-i e \beta \bar{u} \PC{ \frac{i}{2\Delta} F_{2} v_{F}\sigma^{i \nu}q_{\nu} } u  .  \label{S4E3}
\end{equation}
$F_{2}$ in Eq.~(\ref{S4E3}) is the form factor discussed in Sec.~III, and is given by 
\begin{equation}
F_2 (q^{2}\rightarrow 0) = -\frac{\alpha\beta^{3}\bar{R}(\beta)}{2\pi}.  \label{S4E4}
\end{equation}
where
\begin{eqnarray}
\bar{R}(\beta) = \frac{\beta\sqrt{\beta^{2}-1} + \PC{1-6\beta^{2}+4\beta^{4}}\coth^{-1}\PR{\frac{\beta}{\sqrt{\beta^{2}-1} }} }{\beta^{3}\PC{-1+\beta^{2}}^{3/2}}. \nonumber 
\end{eqnarray}
For $\beta \ll 1$ we obtain $\beta^{3} \bar{R}(\beta) \approx -(\pi/2)$, and the correction for the $g_{s}$-factor reads
\begin{eqnarray}
F_2 = \Delta g_{s} = \frac{\alpha}{4}, \label{deltag1}
\end{eqnarray}
whereas for $\beta \approx 1$ (isotropic or fully relativistic limit) the correction is given by
\begin{eqnarray}
 \Delta g_{s} = -\frac{4\alpha}{3\pi}. \label{deltag2}
 \end{eqnarray}
Although $F_{2}=0$ at the tree-level, it acquires a finite value at one-loop. The results ($\ref{deltag1}$) and ($\ref{deltag2}$) show the relevance of using the anisotropic description of PQED. The isotropic model leads to a correction with opposite sign, which decreases the value of the $g$-factor. Besides, the isotropic and the anisotropic theories describe very different physical regimes. 
 
Notice, however, that there is a subtlety in the limit $\beta \to 0$. If one sets $\beta \approx 0$ from the start, the spatial-component contribution to the $S$-matrix element for the scattering from an external field is null (see Eq.~(\ref{2})). This means that there would be no response to an applied external magnetic field. On the other hand, if one keeps $\beta$ and performs the calculations (taking the limit afterwards), as we showed here, one finds a correction to the $g$-factor that is independent of the ratio $v_{F}/c$ between the velocities. This is in agreement with the fact that experiments on the $g$-factor in graphene indicate an enhancement of its bare value $g=2$.

\section{Comparison with experiments}

Even though the gyromagnetic factor is an intrinsic property of the electron in a certain medium, usually it is experimentally determined by applying a magnetic field $B$ perpendicularly to a sample and measuring the Zeeman gap $ \Delta_z = g_{s} \mu_B B$. We have shown in Sec.~III how interaction effects lead to a correction to the bare value $g_s=2$ of the gyromagnetic factor, and we calculated this correction in Sec.~IV. Now, we proceed to compare our theoretical result to the experiments realized in graphene. 

\subsection{Graphene on SiO$_{2}$}

To experimentally probe the enhancement of the gyromagnetic factor due to electron-electron interactions, one needs relatively strong magnetic fields, which lead to orbital quantization. As a result, the enhanced $g$-factor could exhibit a dependence on the Landau-level index $N$ or on the applied $B$-field. In metal-oxide-semiconductors (MOS), this dependence has been theoretically evaluated in Ref.~\cite{Ando-Uemura}, where the authors discuss a theory of oscillatory $g$-factor. This oscillatory behavior has been experimentally observed in GaAs/AlGaAs structures \cite{Nicholas}. Recently, an oscillatory $g$-factor enhancement has been also proposed to occur in the case of graphene at strong magnetic fields \cite{Volkov}. However, measurements of the spin $g$-factor performed by Kurganova et al. for graphene grown on a SiO$_2$ substrate for the different values of the magnetic field, $B=5-7\ {\rm T}$, and Landau levels $N=2 - 10$, did not observe the predicted behavior \cite{Kurganova}. Instead, the authors found that the enhancement of the $g$-factor in graphene in the strong $B$-field regime is \textit{independent} of the Landau level and is \textit{constant} for all extracted data -- exactly as in the case of weak magnetic fields. Their result is compatible with the regime of Gaussian-shaped Landau levels with broadening $\Gamma > g^{*} \mu_B B$ \cite{Kurganova}. Therefore, the computation of the spin splitting within the dynamical electromagnetic interaction performed in Sec.~IV, in the weak-field regime, is appropriate to describe the experiment.  

By evaluating the corrected $g$-factor $g^{*}_{s}$ multiplied by a dimensionless parameter, i. e. by the cyclotron mass $m_c$ in units of the electron mass $m_e$, we obtain the following equivalence
\begin{eqnarray}
\frac{m_c(n) g_{s}^*(n) }{m_e} = g_{s}^*(n) \frac{\hbar\sqrt{\pi n}}{v_F (n) m_e}. \label{function1}
\end{eqnarray}
This expression relates the cyclotron mass $m_c$ to the charge carrier concentration $n$, and to the renormalized Fermi velocity
\begin{equation}
{v}_F(n)=v_F(n_0)\left[1-\frac{\alpha_0 }{8 \varepsilon_G(n)}\ln\left(\frac{n}{n_0}\right)\right]. \label{vren1}
\end{equation}
Here, $\alpha_0 = e^2/ 4\pi\varepsilon_0 \hbar v_F(n_0)$, the vacuum permittivity $\varepsilon_0 = 1$,  and $\varepsilon_G (n)$ is the dielectric constant, which was theoretically and empirically  \cite{Elias} found to depend on the carrier density $n$ (see Ref.~\cite{Mark} for a thorough discussion about the dielectric constant in graphene). 

\begin{figure}[b!]
\centering
\includegraphics[width=0.95\columnwidth]{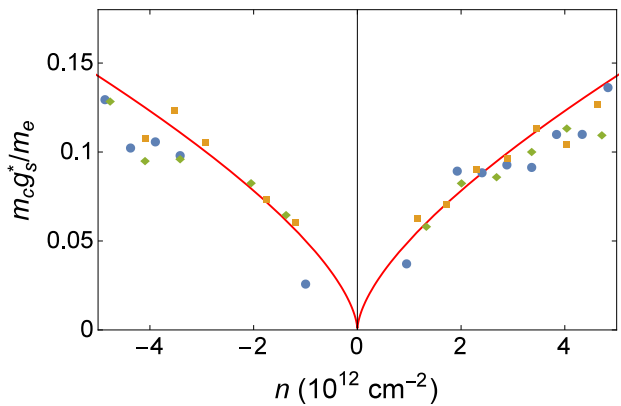}  
\caption{\textbf{$g_s$-factor enhanced due to electron-electron interactions}. At high densities, the theoretical red curve is given by Eq.~(\ref{deltag1}), together with the renormalized value of $v_F (n)$ given by Eq.~(\ref{vren1}), and the reference value $v_F^0 =1\times 10^6 $ m/s. Here, $\alpha=0.9$ (i.e. $\varepsilon_{G}=2.44$), which is the bare fine structure constant for graphene on SiO$_{2}$ \cite{Mark}. 
\label{Fig1a}} 
\end{figure}

It is known that the logarithm in the renormalized Fermi velocity $v_{F}$ in graphene arises due to electron-electron interactions. For \textit{undoped} graphene, via renormalization-group methods one finds that $v_{F}$ depends on the smallest energy scale of the theory at which the RG flow is suppressed, namely the doping energy $\propto n$. If one considers \textit{doped} graphene, this logarithmic dependence is not altered \cite{DasSarma2}, but the effective interaction parameter is modified, i.e. $\alpha \rightarrow \alpha^{*}$. 
We have accounted for this effect by considering a dielectric function that depends on $n$.

 \begin{figure}[b!]
\centering  
\includegraphics[width=0.95\columnwidth]{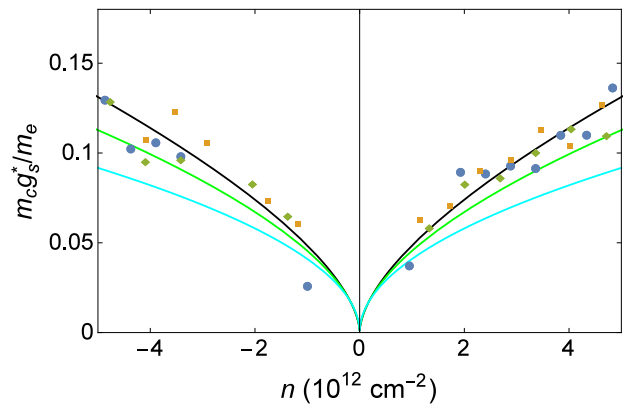}
\caption{\textbf{Dependence of the $g_s$-factor on the dielectric constant $\varepsilon_G$}. The black and green solid curves correspond to different values of the dielectric constant, chosen {\it ad hoc} to be $\varepsilon_G = 3$ and 5, respectively. The light-blue solid curve denotes the bare $g_{s}=2$ factor \cite{Kurganova}. All the theoretical curves are given by Eq.~(\ref{deltag1}), together with the renormalized value of $v_F (n)$ given by Eq.~(\ref{vren1}), and the reference value $v_F^0 =1\times 10^6 $ m/s.
\label{Fig1b}} 
\end{figure}

The parameter $g_{s}^*$ in Eq.~(\ref{function1}) is the effective $g_{s}$-factor, which, in the experimental work, is taken to be the constant parameter that best fits the experimental points \cite{Kurganova}. Recalling that the bare $g_{s}$-factor in graphene is $g_{s} = 2$, and replacing Eqs.~(\ref{deltag1}) and (\ref{vren1}) into Eq.~(\ref{function1}), we obtain the corrected $g_{s}$-factor $g_{s}^*=2+2\Delta g_s$, 
\begin{eqnarray}
&&\frac{m_c (n) g_{s}^*(n)}{m_e}= \PC{2+\frac{\alpha}{2}}\frac{\hbar\sqrt{\pi n}}{ m_e v_F^0 }\frac{1}{ \PR{1+\frac{\alpha_0}{8\varepsilon_G}\ln\PC{\frac{n_0}{n}}}}  \nonumber \\
&&= \frac{2}{ \PR{1+\frac{\alpha_0}{8\varepsilon_G}\ln\PC{\frac{n_0}{n}}}}\frac{\hbar\sqrt{\pi n}}{ m_e v_F^0 }\nonumber \\
&&+\frac{e^{2}}{8\pi\hbar\varepsilon_{0} \varepsilon_G v_{F}^{0}\PR{1+\frac{\alpha_0}{8\varepsilon_G}\ln\PC{\frac{n_0}{n}}}^{2}} \frac{\hbar\sqrt{\pi n}}{ m_e v_F^0 },
\end{eqnarray}
where we used $\alpha=e^2/(4\pi \varepsilon_0 \varepsilon_G \hbar v_F(n))$, with $v_F(n)$ the renormalized Fermi velocity given by Eq.~(\ref{vren1}). Note that screening is taken into account in $\alpha$ and in $v_F(n)$. Choosing the reference value of $n_0$ around the values of $n$ that we want to describe, and neglecting corrections of order $(\alpha_{0}/\varepsilon_G)^{2}$, we may write
\begin{eqnarray}
\PR{1+\frac{\alpha_0}{8\varepsilon_G}\ln\PC{\frac{n_0}{n}}}^{2} &\approx& 1 + \frac{2\alpha_0}{8\varepsilon_G}\ln\PC{\frac{n_0}{n}}  \nonumber
\end{eqnarray}
to obtain
\begin{eqnarray}
\frac{m_c g_{s}^*}{m_e} &=&\left\{ \frac{2}{ \PR{1+\frac{\alpha_0}{8\varepsilon_G}\ln\PC{\frac{n_0}{n}}}}
\right. \nonumber \\
&+&\left. \frac{\alpha_0}{2 \varepsilon_G \PR{1 + \frac{2\alpha_0}{8\varepsilon_G}\ln\PC{\frac{n_0}{n}}}}    \right\} \frac{\hbar\sqrt{\pi n}}{ m_e v_F^0 }. \label{y3n}  
\end{eqnarray}

In Fig.~\ref{Fig1a}, we plot Eq.~(\ref{y3n}) for the value of $\alpha^{*}_{0}=\alpha_0/\varepsilon_G=0.9$ (i.e. $\varepsilon_{G}=2.44$), as given in Ref.~\cite{Mark} for graphene on SiO$_{2}$ \cite{footnote}. The theoretical curve exhibits a very good agreement with the experimental data, indicating that interaction effects are able to capture the behavior of the $g$-factor in this material. This is the main result of this subsection. Notice that there are \textit{no} fitting parameters in Fig.~\ref{Fig1a}.

We proceed by investigating how the parameters in the theory, such as dielectric constant $\varepsilon_G$ and bare Fermi velocity $v_{F}^{0}$, modify the curve obtained in Fig.~\ref{Fig1a}. For {\it ad hoc} values of the dielectric constant $\varepsilon_G = 3$ (black) and 5 (green), we plot Eq.~(\ref{y3n}) in Fig.~\ref{Fig1b}. Upon increasing $\varepsilon_G$, the curve bends down for large carrier concentration values. The light-blue curve, corresponding to the bare value of the $g$-factor $g_s = 2$ clearly cannot describe the observed data, thus confirming the relevance of interactions in the description of the spin $g$-factor.

\begin{figure}[tbh]
\centering
\includegraphics[width=0.95\columnwidth]{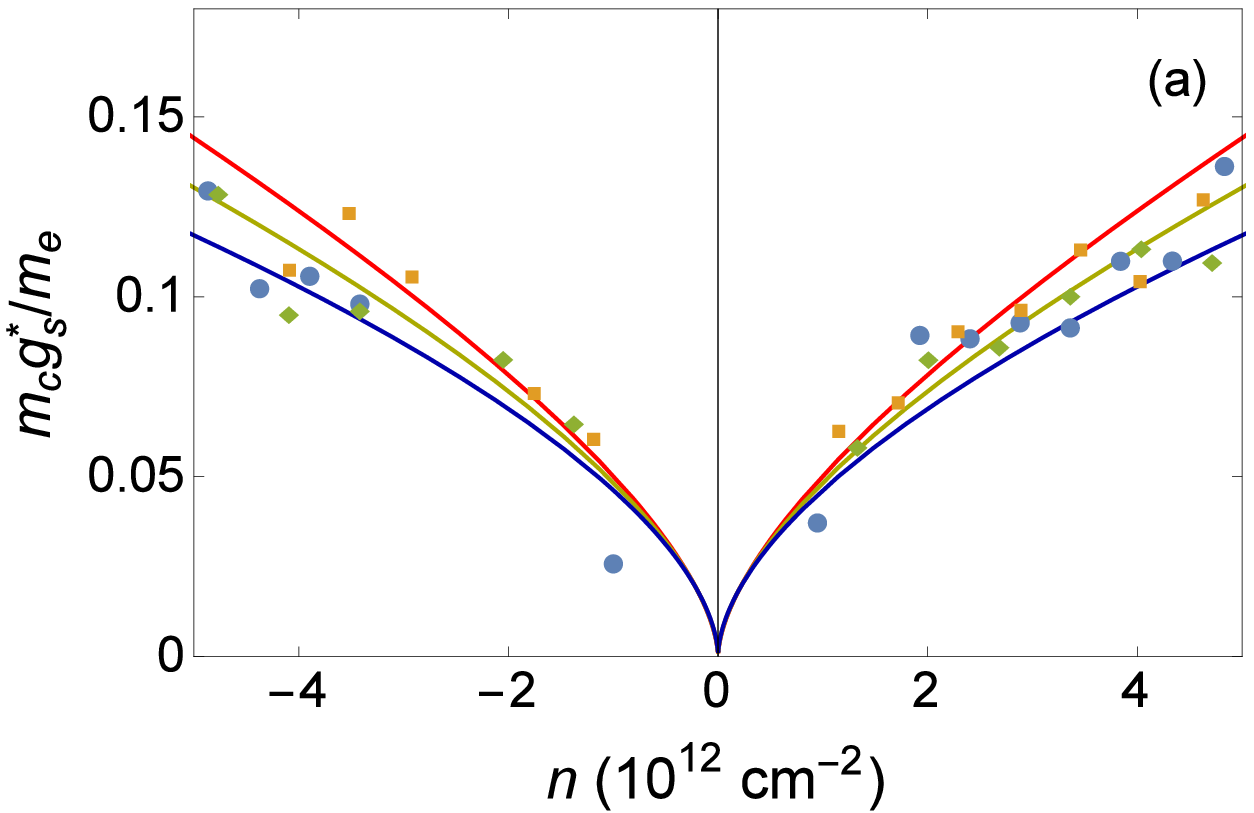} \qquad \quad
\includegraphics[width=0.92\columnwidth]{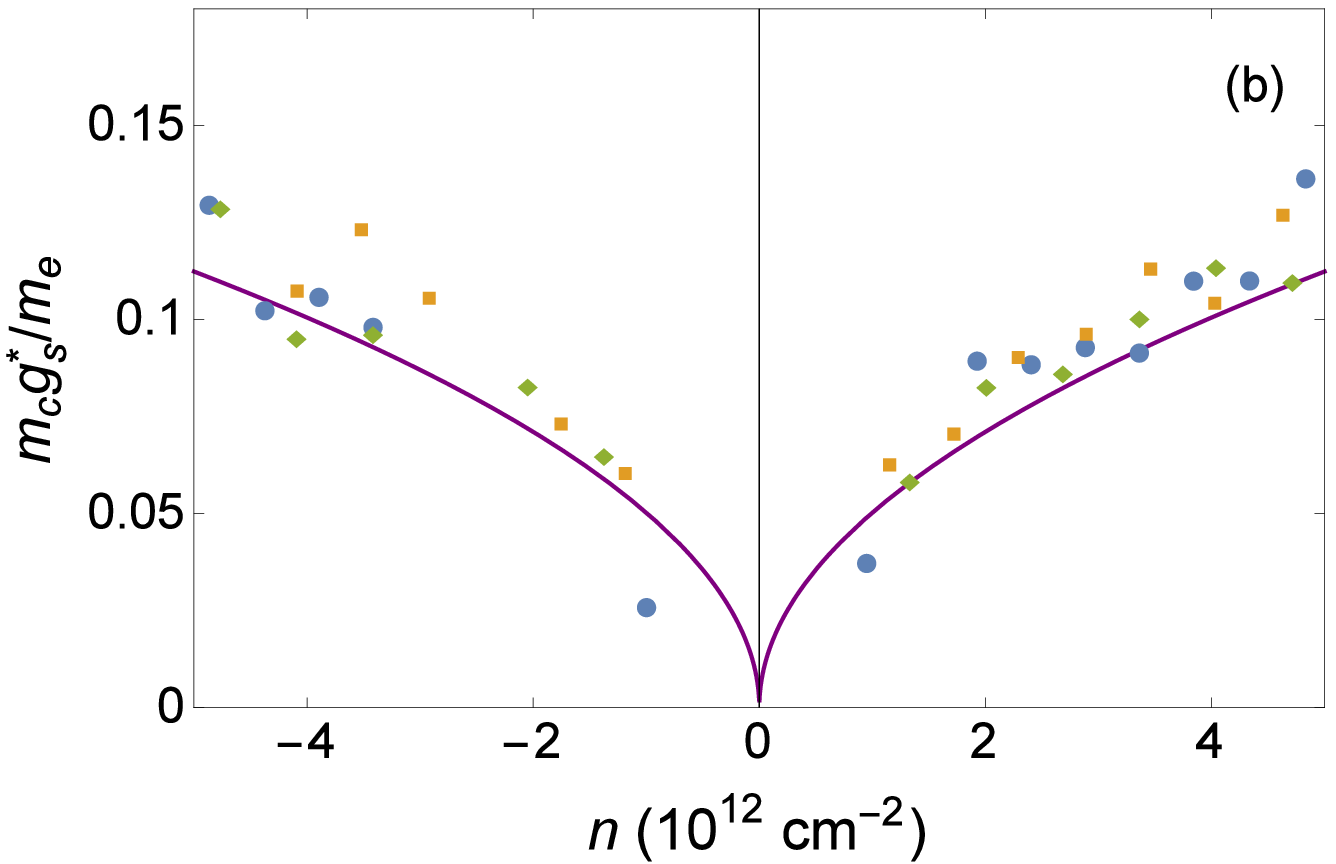}
\caption{\textbf{Dependence of the $g_s$-factor on the reference value $v_F^0$}. (a) The red curve is the same as in Fig.~\ref{Fig1a}, for $v_F^0=1\times10^6{\rm m/s}$, the yellow and blue curves are given by Eq.~(\ref{y3n}) with $v_F^0=1.25\times10^6{\rm m/s}$ and $v_F^0=1.75\times10^6{\rm m/s}$, respectively. We use $\varepsilon_G=2.44$ for the three curves. 
(b) The purple curve is obtained from Eq.~(\ref{deltag1}) for a non-renormalized $v_F^0=1\times10^6{\rm m/s}$ and $\varepsilon_G=2.44$, which results in a spin $g$-factor $g_s^{*}\approx 2.45$. \label{Fig2}}
\end{figure}

After having verified the trend of the $g_s$-factor renormalization upon varying the dielectric constant $\varepsilon_G$, as shown in Fig.~\ref{Fig1b}, we compare the behavior of $g_s^{*}$ upon fixing $\varepsilon_G$ and varying the reference point $v_F^0$, which arises within the RG procedure. The dependence on $v_{F}^{0}$ may be observed in Fig.~\ref{Fig2} (a), for the range of values compatible with the findings of Ref.~\cite{Elias}. 

To complete the analysis, we also compare the value expected for the renormalization of the $g_s$-factor for the case of a non-renormalized Fermi velocity. In this case, by using a dielectric constant $\varepsilon_G=2.44$, we obtain the value $g_s^{*}\approx 2.45$, which is represented by the purple curve in Fig.~\ref{Fig2} (b). We can clearly observe the difference between the curves of Figs.~\ref{Fig1a} and \ref{Fig2} (b), where in the first we used a renormalized Fermi velocity, while in the second not.

\subsection{Graphene on SiC(111)}

Measurements of the spin $g$-factor were performed also in graphene on SiC \cite{Stroscio}, where the top layer of multilayer epitaxial graphene grown on SiC was investigated by high-resolution scanning tunneling spectroscopy. At ultra-low temperatures, in extremely clean samples, these spin degeneracies may be lifted and the authors reported a small correction to the bare spin $g$-factor $\Delta g^*_s \approx 0.23 - 0.36$. These values $g^*_{s,K} = 2.23$ and $g^*_{s,K'} = 2.36$ (there is a small difference in the value measured for each of the valleys) are also comparable to the one obtained by Kurganova et al.  \cite{Kurganova} for graphene grown on SiO$_2$,  $g^*_s = 2.7 \pm 0.2$. 

\begin{figure}[t!]
\centering
\includegraphics[width=0.90\columnwidth]{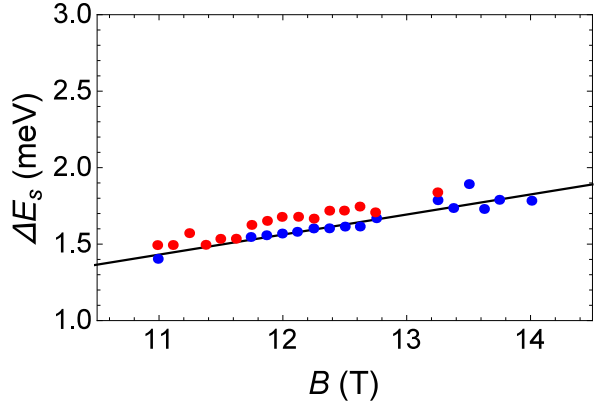}   
\caption{{\bf Spin $g$-factor in graphene grown on SiC}. Comparison between theory and experiments for the spin $g$-factor. In the experiments, there is an asymmetry between the valleys, indicated by the red and blue points. They lead to a spin $g$-factor of $g^*_{s,K} = 2.23 \pm 0.01$ and $g^*_{s,K'} = 2.36 \pm 0.01$, respectively \cite{Stroscio}. The black-solid line, which provides a good agreement with the experimental data, is obtained by using Eq.~(\ref{deltag1}) and the fitting parameter $\alpha_{0}^{*} = 0.51$, since the precise value of the dielectric constant is unknown. The reference value for the magnetic field in the RG equations for the renormalized Fermi velocity used here is $B_0 = 14$ T.  
\label{GsStroscio} }
\end{figure}

We now confront these data to our results obtained within the PQED. In this experiment, the Zeeman splitting was measured, which is given by
\begin{eqnarray}
\Delta E_{s} = g^{*}_{s} (B) \mu_{B} B.
\end{eqnarray}
Inserting the value found for $g^{*}_{s}= 2+2\Delta g_s$ with $\Delta g_s$ given by Eqs.~(\ref{deltag1}) and (\ref{vren1}), we obtain \cite{vFB}
\begin{eqnarray}
\Delta E_{s} = \chav{2+\frac{\alpha_{0}}{2\varepsilon_{G}\PR{1+\frac{\alpha_{0}}{8\varepsilon_{G}}\ln\PC{\frac{B_0}{B}}}}} \mu_{B} B. \label{spinsp}
\end{eqnarray}
We can observe in Ref.~\cite{Stroscio} that the experimentally detected spin-splitting does not change much when increasing the magnetic field from 11 to 14 T. We plot Eq.~(\ref{spinsp}) for the spin-splitting in Fig.~\ref{GsStroscio} using $\varepsilon_G$ as a fitting parameter. By using $v_{F}^{0}=1.08\times 10^{6}$ m/s \cite{Stroscio}, we find that $\varepsilon_G\approx 4$ for this sample, which falls within the range of values discussed in Ref.~\cite{Siegel} for monolayer graphene on SiC. 

\section{Conclusions}

In this work, we have investigated the corrections to the spin gyromagnetic factor in graphene that are generated due to electronic interactions. The calculations were performed in the framework of the anisotropic pseudo-QED, which is a theory that takes into account the full electromagnetic interaction and breaks Lorentz symmetry by considering two different velocities: $c$ for the photons and $v_F$ for the electrons. With these two ingredients, we have obtained an explicit expression for the spin $g$-factor correction, which has allowed us to compare our theoretical findings with experiments on graphene deposited on SiO$_2$ and on SiC. 

The outcome of the comparison indicates that the renormalization of the Fermi velocity is very important to better describe the experiments. By combining this renormalization effect and choosing the dielectric constant according to the substrate, we have shown in Fig.~\ref{Fig1a} a very good agreement between our theoretical results and the experimental data. 

Our work confirms the importance of electronic interactions in the description of graphene, and indicates that the pseudo-QED formalism is able to capture its signatures in great detail.

\acknowledgments
This work was supported by CNPq (Brazil), CAPES (Brazil), FAPERJ (Brazil), NWO-VICI (Netherlands) and by the Brazilian government project Science Without Borders. We are grateful to D. Haldane, S. Sachdev, D. V. Khveshchenko, A. H. Castro Neto,  L. Fritz, M. Goerbig, and V. Juricic for the discussions. V.S.A. acknowledges  NWO and the Institute for Theoretical Physics of Utrecht University for the kind hospitality.\\

\appendix
\begin{widetext}
\section{Form factor detailed calculation}

In this appendix, we present the details of the calculation of Sec.~IV. By using the anticommutation of the gamma matrices and the Dirac equations in momentum space $\bar{u}(\bar{p}') \displaystyle{\not}\bar{p}'=   \bar{u}(\bar{p}') \Delta$ and  $ \displaystyle{\not} \bar{p} u(\bar{p})  =  \Delta u(\bar{p})$, we can rewrite Eq.~(\ref{S4E2}) of the main text as 
\begin{eqnarray}
i\Omega^{i}&=&  \frac{3e^3v_F}{8\varepsilon}\int_{0}^{1}dx\int_{0}^{1-x}dy (1-x-y)^{-1/2} \int \frac{d^{2}k}{(2\pi)^{2}}\int_{-\infty}^{\infty}\frac{dk_{0}}{(2\pi)}  \frac{\bar{u}(\bar{p}')[k_{0}^{2}\gamma^{i} + V^i+N^i]u(\bar{p})}{\PC{k_{0}^{2} - \Lambda}^{5/2}}. \label{App1}
\end{eqnarray}

In the equation above, we used the parametric integral
\begin{eqnarray}
\chav{\PR{(\bar{k}+\bar{p}')^{2}-\Delta^{2}}\PR{(\bar{k}+\bar{p})^{2}-\Delta^{2}} \sqrt{k^{2}}}^{-1}=\frac{3}{4}\int_{0}^{1}dx\int_{0}^{1-x}dy \frac{(1-x-y)^{-1/2}}{\PR{(k_{0}+\omega_{0})^{2} - \Lambda}^{5/2}}, \nonumber
\end{eqnarray}
where
$$\Lambda = - A\PC{\textbf{k}^{2} -v_{F}^{2}A^{-1}\boldsymbol{\omega}}^{2}+ \omega_{0}^{2} + A^{-1}v_{F}^{4} \boldsymbol{\omega}^{2},$$
with $A=\PR{- v_{F}^{2}(x+y) -c^{2}(1-x-y)}$, $\omega_{0}=(p'_{0}x+p_{0}y)$ and $\boldsymbol{\omega}=(\textbf{p}'x+ \textbf{p}y)$. We performed also the displacement $k_0\rightarrow k_0-\omega_0$, such that the terms in the numerator of Eq.~(\ref{App1}) become

\begin{eqnarray}
V^i &\rightarrow & [\omega_{0}^{2} -2\omega_{0}(p'_{0}+ p_{0}) + 4p'_{0}p_{0}]\gamma^{i} + 2\gamma^{i}\gamma_{0}p'^{0}v_{F}\boldsymbol{\gamma}\cdot\textbf{k}+ 2p_{0}v_{F}\gamma^{0}\gamma^{l}k_{l}\gamma^{i} -  2v_{F}k^{i} \omega_{0}\gamma^{0} + \nonumber \\
&-&  2v_{F}^{2}k^{i}k_{l}\gamma^{l} +(1/2)v_{F}^{2}\textbf{k}^{2}\gamma^{i} \nonumber
\end{eqnarray}
and 
\begin{eqnarray}
N^{i} \beta^{-2} &\rightarrow & - 4v_F^{2}\gamma^{i}\chav{(1-v_{F})\textbf{k}^{2}+ \textbf{k}\cdot(\textbf{p}'+\textbf{p}) + \textbf{p}'\cdot \textbf{p}} +4v_Fp'^{i}\gamma^{0} \omega_{0} + 4v_Fk^{i}\gamma^{0}\omega_{0} -2v_F(p'_j+p_{j})\gamma^{j}\gamma^{i}\gamma^{0}\omega_{0}+ \nonumber \\
&+&2v_F(p_{0}-p'_{0})\gamma^{0}\boldsymbol{\gamma} \cdot\textbf{k}\gamma^{i}+ 4v_F(\Delta-\gamma^{0}p'_{0})k^{i} + 4v_{F}^{2}(p^{i}+ p'^{i})\boldsymbol{\gamma}\cdot \textbf{k} , \nonumber 
\end{eqnarray}
where we eliminated the odd terms in $k_{0}$. 

As a next step, we try to simplify the lengthy expressions. Since we are interested in obtaining the gyromagnetic factor, we will disregard the terms proportional to $\gamma^i$. After solving the integral over $k_0$, we find
$$i\Omega_{gy}^{i}=  \frac{3e^3v_F}{16\pi\varepsilon}\int_{0}^{1}dx\int_{0}^{1-x}dy (1-x-y)^{-1/2} \int \frac{d^{2}{\bf k}}{(2\pi)^{2}}\frac{\bar{u}(\bar{p}')[\frac{4}{3}(-2 v_F^3 A^{-1} \omega^i \omega_0\gamma^0+2 v_F^6 A^{-2} \omega^i \boldsymbol{\gamma}\cdot\boldsymbol{\omega})+\beta^2 L^i]u(\bar{p})}{A^2\PR{(\textbf{k}^{2}-v_F^2 A^{-1}\boldsymbol{\omega})^2 - \tilde{\Lambda}}^{5/2}},$$
where $L^i= 4\Delta v_{F}^{3}A^{-1}\omega^{i} + 4v_{F}^{4}A^{-1}(p^{i}+ p'^{i})\boldsymbol{\gamma}\cdot \boldsymbol{\omega} $ and 
$\tilde{\Lambda}=  \PC{\omega_{0}^{2} + A^{-1}v_{F}^{4} \boldsymbol{\omega}^{2}}A^{-1}. $
 Displacing $\textbf{k}\rightarrow \textbf{k}+v_F^2 A^{-1}\boldsymbol{\omega}$, we find, after solving the integrals over $\textbf{k}$, 
$$i\Omega^{i}_{gy}=-\frac{e^{3}v_{F}}{16\pi^{2}\varepsilon}\int_{0}^{1}dx\int_{0}^{1-x}dy (1-x-y)^{-1/2} \PC{\frac{2v_F^6 A^{-2} w^i \boldsymbol{\gamma}\cdot \boldsymbol{\omega}+\beta^{2}L^i}{A^{2}\tilde{\Lambda}}}, $$
where we considered $p_{0}=p'_{0}=0$. Therefore, working on mass-shell, we can use that $v_f\gamma^jp_j=\Delta$. By using that $2p'^{i}=P^{i}+q^{i}$ and $2p^{i}= P^{i}-q^{i}$, we can write
$2v_F^6 A^{-2} w^i \boldsymbol{\gamma}\cdot \boldsymbol{\omega}\rightarrow - \Delta v_F^5 A^{-2} P^i (x+y)^2 $
and
$L^i\rightarrow -2 \Delta v_F^3 A^{-1} P^i (x+y).$
Now, we can use the Gordon identity
$$\bar{u}P^{i}u = 2\Delta\bar{u}\gamma^{i}u - i\bar{u}\sigma^{i\nu}q_{\nu}u,$$
and write
$$\Omega^{i}_{gy}=-i e \beta \bar{u} \PC{ \frac{i}{2\Delta} F_{2} v_{F}\sigma^{i \nu}q_{\nu} } u  . $$
Hence, the form factor $F_2$ is identified as
\begin{equation}
F_2 = -\frac{\alpha\beta }{2\pi}\int_{0}^{1}dx\int_{0}^{1-x}dy (1-x-y)^{-1/2} \chav{\frac{ \Delta^{2}v_{F}^{2}\PR{2\PC{x+y}-\frac{\PC{x+y}^{2}}{\beta^{2}(x+y)+(1-x-y)}} }{Av_{F}^{2}(\textbf{p}'x+\textbf{p}y)^{2}}}.  \label{App2}
\end{equation}
By rewriting the denominator of Eq.~(\ref{App2}) as
$$v_{F}^{2}(\textbf{p}'x+\textbf{p}y)^{2}=-\Delta^{2}(x+y)^{2} + q^{2}xy,$$
with $q^{2}=(p'-p)^{2}$ and using that $q^2\rightarrow 0$, we obtain
\begin{equation}
F_2 =-\frac{\alpha\beta^{3}\bar{R}(\beta)}{2\pi}, \label{App3}
\end{equation}
with
\begin{equation}
\bar{R}(\beta)= \int_{0}^{1}dx\int_{0}^{1-x}dy  \frac{2(1-x-y)^{-1/2}}{(x+y)[\beta^{2}(x+y)+(1-x-y)]} - \int_{0}^{1}dx\int_{0}^{1-x}dy  \frac{(1-x-y)^{-1/2}}{[\beta^{2}(x+y)+(1-x-y)]^{2}}. \label{App4}
\end{equation}
In the limit of $v_{F}=c=1$, we find 
\begin{eqnarray}
\int_{0}^{1}dx\int_{0}^{1-x}dy   \frac{(1-x-y)^{-1/2}\PC{2-x-y}}{(x+y)} = \frac{8}{3},\nonumber
\end{eqnarray}
which is exactly what is obtained in the isotropic model. 

On the other hand, if we solve the integrals in Eq.~(\ref{App4}), we find
\begin{eqnarray}
\bar{R}(\beta) = \frac{\beta\sqrt{\beta^{2}-1} + \PC{1-6\beta^{2}+4\beta^{4}}\coth^{-1}\PR{\frac{\beta}{\sqrt{\beta^{2}-1} }} }{\beta^{3}\PC{-1+\beta^{2}}^{3/2}}.
\end{eqnarray}

\end{widetext}


\begin{thebibliography}{100}

\bibitem{Schwinger48} J. Schwinger, Phys. Rev. {\bf 73}, 416 (1948).

\bibitem{expvan}R. S. Van Dyck Jr., P. B. Schwinberg and H. G. Dehmelt, Phys. Rev. Lett. {\bf 59}, 26 (1987).

\bibitem{Hanneke} D. Hanneke, S. Fogwell and G. Gabrielse, Phys. Rev. Lett. {\bf 100}, 120801 (2008).

\bibitem{Kinoshita} T. Aoyama, M. Hayakawa, T. Kinoshita and M. Nio, Phys. Rev. Lett. {\bf 109}, 111807 (2012).

\bibitem{Kovner} A. Kovner and B. Rosenstein, Phys. Rev. B {\bf 42}, 4748 (1990); N. Dorey and N. E. Mavromatos,  Nucl. Phys. B {\bf 386},  614 (1992); S. Teber, Phys. Rev. D \textbf{86}, 025005 (2012); S. Teber, Phys. Rev. D \textbf{89}, 067702 (2014).

\bibitem{Marino} E. C. Marino, Nucl. Phys. B {\bf 408}, 551 (1993).

\bibitem{Geim1} K. S. Novoselov \textit{et al.}, Nature \textbf{438}, 197 (2005).

\bibitem{Castro} A. H. Castro Neto, F. Guinea, N. M. R. Peres, K. S. Novoselov and A. K. Geim, Rev. Mod. Phys. \textbf{81}, 109 (2009).

\bibitem{Eva} X. Du, I. Skachko, F. Duerr, A. Luican and E. Y. Andrei, Nature \textbf{462}, 192 (2009).

\bibitem{Kim1} K. I. Bolotin, F. Ghahari, M. D. Shulman, H. L. Stormer and P. Kim, Nature {\bf 462}, 196 (2009).

\bibitem{Kim2} C. R. Dean \textit{et al.}, Nat. Phys. {\bf 7}, 693 (2011).

\bibitem{Luican} A. Luican, G. Li and E. Y. Andrei, Phys. Rev. B {\bf 83}, 041405(R) (2011). 

\bibitem{Elias} D. C. Elias \textit{et al.}, Nat. Phys. {\bf 7}, 701 (2011).

\bibitem{Stroscio2} J. Chae \textit{et al.}, Phys. Rev. Lett. {\bf 109}, 116802 (2012).

\bibitem{Maria2} J. Gonz\'{a}lez, F. Guinea and M. A. H. Vozmediano, Nucl. Phys. B {\bf 424}, 595 (1994); Phys. Rev. B {\bf 59}, 2474 (1999). 

\bibitem{Foster} M. S. Foster and I. L. Aleiner, Phys. Rev. B {\bf 77}, 195413 (2008).

\bibitem{Maria3} J. Gonz\'{a}lez, F. Guinea and M. A. H. Vozmediano, Phys. Rev. Lett. {\bf 77}, 3589 (1996).

\bibitem{Maria0} F. Juan, A. G. Grushin and M. A. H. Vozmediano, Phys. Rev. B {\bf 82}, 125409 (2010). 

\bibitem{Maria} M. A. H Vozmediano and F. Guinea, Phys, Scr. {\bf T146}, 014015 (2012).

\bibitem{dasSarma} E. Barnes, E. H. Hwang, R. E. Throckmorton and S. Das Sarma, Phys. Rev. B { \bf 89}, 235431 (2014).

\bibitem{us} E. C. Marino, L. O. Nascimento, V. S. Alves and C. Morais Smith, Phys. Rev. X {\bf 5}, 011040 (2015).

\bibitem{Miransky} E. V. Gorbar, V. P. Gusynin, V. A. Miransky and I. A. Shovkovy, Phys. Scr. \textbf{T146}, 014018 (2012); E. V. Gorbar, V. P. Gusynin and V. A. Miransky, Phys Rev D \textbf{64}, 105028 (2001).

\bibitem{marinorubens} R. L. P. G. Amaral and E. C. Marino, J. Phys. A {\bf 25}, 5183 (1992).

\bibitem{us2} E. C. Marino, L. O. Nascimento, V. S. Alves and C. Morais Smith, Phys. Rev. D {\bf 90}, 105003 (2014).

\bibitem{Kurganova} E. V. Kurganova \textit{et al.}, Phys. Rev. B {\bf 84}, 121407 (2011).

\bibitem{Stroscio} Y. J. Song \textit{et al.}, Nature {\bf 467}, 185 (2010). 

\bibitem{Ando} M. Koshino and T. Ando, Phys. Rev. B {\bf 81}, 195431 (2010).

\bibitem{Hooft} C. G Bollini and J. J. Giambiagi, Phys. Lett. B {\bf 40}, 566 (1972); G.'t Hooft and M. Veltman, Nucl. Phys. B {\bf 44}, 189 (1972).

\bibitem{Peskin} M. E. Peskin and D. V. Schroeder, An Introduction to Quantum Field Theory, Addison-Wesley Publishing Company (1995).

\bibitem{Drell} J. D. Bjorken and S. D. Drell, Relativistic Quantum Mechanics, McGraw-Hill Book Company (1964).

\bibitem{Ando-Uemura} T. Ando and Y. Uemura, J. Phys. Soc. Jpn. \textbf{37}, 1044 (1974).

\bibitem{Nicholas} R. J. Nicholas, R. J. Haug, K. v. Klitzing and G. Weimann, Phys. Rev. B \textbf{37}, 1294 (1988).

\bibitem{Volkov} A. V. Volkov, A. A. Shylau and I. V. Zozoulenko, Phys. Rev. B {\bf 86}, 155440 (2012).

\bibitem{Mark} M. Goerbig, Rev. Mod. Phys. {\bf 83}, 1193 (2011).

\bibitem{DasSarma2} S. Das Sarma, E. H. Hwang and Wang-Kong Tse, Phys. Rev. B \textbf{75}, 121406(R) (2007).

\bibitem{footnote} It was shown in Ref.~\cite{Elias} that it is possible to describe the renormalization of the Fermi velocity in leading order, given that a more elaborated RPA expression is used for the dielectric constant $\varepsilon_G$ because it depends on the doping concentration $n$. Although the changes in $\varepsilon_G$ upon varying $n$ from $10^{9}$ to $10^{12}$ cm$^{-2}$ would be significant, in the regime of the experimental data \cite{Kurganova} used in our manuscript the carrier concentration $n \approx 1-4 \times 10^{12}$ cm$^{-2}$ remains nearly constant.

\bibitem{vFB} Shizuya, K., Phys. Rev. B \textbf{81}, 075407 (2010).

\bibitem{Siegel} D. A. Siegel \textit{et al.}, Proc. Natl. Acad. Sci. USA \textbf{108}, 11365 (2011).




    

\end{thebibliography}
\end{document}